\documentclass{birkmult}

\newcommand{\SetFigFont}[3]{}

\newtheorem{Def}{Def.}[section]
\newtheorem{Thm}[Def]{Theorem}

\newcommand{\spc}{\;\;\;\;\;\;\;\;\;\;}

\newcommand{\bra}{\mbox{$< \!\!$ \nolinebreak}}
\newcommand{\ket}{\mbox{\nolinebreak $>$}}

\newcommand{\R}{\mathbb{R}}
\newcommand{\1}{\mbox{\rm 1 \hspace{-1.05 em} 1}}

\newcommand{\Pdd}{\mbox{$\partial$ \hspace{-1.2 em} $/$}}
\newcommand{\slsh}{\mbox{ \hspace{-1.1 em} $/$}}
\newcommand{\Tr}{\mbox{\rm{Tr}\/}}
\newcommand{\beq}{\begin{equation}}
\newcommand{\eeq}{\end{equation}}

\newcommand{\Aslsh}{\mbox{ $\!\!A$ \hspace{-1.2 em} $/$}}

\begin{document}
%
\firstpage{1}
%
%
%
%
%
%
%
%
\title{The Principle of the Fermionic Projector: \\
An Approach for Quantum Gravity?}
\author{Felix Finster}

\address{%
NWF I -- Mathematik \\
Universit\"at Regensburg \\
D-93040 Regensburg, Germany}

\email{Felix.Finster@mathematik.uni-regensburg.de}

\thanks{I would like to thank the Erwin Schr\"odinger Institute,
Wien, for its hospitality.}

\markboth{Felix Finster}{The Principle of the Fermionic Projector}

\maketitle
\begin{abstract}
In this short article we introduce the mathematical framework of the principle
of the fermionic projector and set up a variational principle in discrete
space-time. The underlying physical principles are discussed.
We outline the connection to the continuum theory and state
recent results. In the last two sections, we
speculate on how it might be possible to describe
quantum gravity within this framework.
\end{abstract}

The principle of the fermionic projector~\cite{PFP} provides a new model of
space-time together with the mathematical framework for the formulation of
physical theories. It was proposed to formulate physics in this framework
based on a particular variational principle. Here we
explain a few basic ideas of the approach, report on recent results
and explain the possible connection to quantum gravity.

It is generally believed that the concept of a space-time continuum (like
Minkowski space or a Lorentzian manifold) should be modified for distances as
small as the Planck length. We here assume that space-time is discrete on the
Planck scale. Our notion of ``discrete space-time'' differs from other discrete
approaches (like for example lattice gauge theories or quantum foam models) in
that we do not assume any structures or relations between the space-time
points (like for example the nearest-neighbor relation on a space-time lattice).
Instead, we set up a variational principle for an ensemble of quantum mechanical
wave functions. The idea is that for mimimizers of our variational principle,
these wave functions should induce relations between the discrete space-time points, which, in a suitable limit, should go over to the topological and causal structure of a Lorentzian manifold. More specifically, in this limit the wave functions should group to a configuration of Dirac seas.

For clarity, we first introduce the mathematical framework (Section~\ref{sec1}) and discuss it afterwards, working out the underlying physical
principles (Section~\ref{sec2}). Then we outline the connection to the continuum
theory (Sections~\ref{sec3} and~\ref{sec4}) and state
some results (Sections~\ref{sec5}). Finally, we give an outlook on
classical gravity (Section~\ref{sec6}) and
the field quantization (Section~\ref{sec7}).

\section{A Variational Principle in Discrete Space-Time} \label{sec1}
\markboth{Felix Finster}{The Principle of the Fermionic Projector}
We let~$(H, \bra .|. \ket)$ be a complex inner product space of signature~$(N,N)$.
Thus~$\bra .|. \ket$ is linear
in its second and antilinear in its first argument, and it is symmetric,
\[ \overline{\bra \Psi \:|\: \Phi \ket} \;=\; \bra \Phi \:|\: \Psi \ket \quad
\spc {\mbox{for all~$\Psi,\Phi \in H$}} \,, \]
and non-degenerate,
\[ \bra \Psi \:|\: \Phi \ket \;=\; 0 \;\;\; {\mbox{for all $\Phi \in H$}}
 \quad \Longrightarrow \quad
\Psi \;=\; 0 \:. \]
In contrast to a scalar product, $\bra .|. \ket$ is {\em{not}} positive.
Instead, we can choose an orthogonal
basis~$(e_i)_{i=1,\ldots,2N}$ of~$H$ such that the inner product
$\bra e_i \,|\, e_i \ket$ equals~$+1$ if $i=1,\ldots,N$ and equals~$-1$ if
$i=N+1,\ldots,2N$.

A {\em{projector}}~$A$ in~$H$ is defined just as in Hilbert spaces as a linear
operator which is idempotent and self-adjoint,
\[ A^2 = A \spc {\mbox{and}} \spc \bra A\Psi \:|\: \Phi \ket = \bra \Psi \:|\: A\Phi \ket \quad
{\mbox{for all $\Psi, \Phi \in H$}}\:. \]
Let~$M$ be a finite set. To every point~$x \in M$ we associate a projector
$E_x$. We assume that these projectors are orthogonal and
complete in the sense that
\beq \label{oc}
E_x\:E_y \;=\; \delta_{xy}\:E_x \spc {\mbox{and}} \spc
\sum_{x \in M} E_x \;=\; \1\:.
\eeq
Furthermore, we assume that the images~$E_x(H) \subset H$ of these
projectors are non-de\-ge\-ne\-rate subspaces of~$H$, which
all have the same signature~$(n,n)$.
We refer to~$(n,n)$ as the {\em{spin dimension}}.
The points~$x \in M$ are
called {\em{discrete space-time points}}, and the corresponding
projectors~$E_x$ are the {\em{space-time projectors}}. The
structure~$(H, \bra .|. \ket, (E_x)_{x \in M})$ is
called {\em{discrete space-time}}.

We introduce one more projector~$P$ in~$H$, the so-called
{\em{fermionic projector}}, which has the additional property that its
image~$P(H)$ is a {\em{negative definite}} subspace of~$H$.
We refer to the rank of~$P$ as the {\em{number of particles}}
$f := \dim P(H)$.

A space-time projector~$E_x$ can be used to project vectors of~$H$
to the subspace~$E_x(H) \subset H$. Using a more graphic notion, we
also refer to this projection as the {\em{localization}} at the
space-time point~$x$.
For example, using the completeness of the space-time projectors~(\ref{oc}),
we readily see that
\beq \label{loctr}
f \;=\; \Tr \,P \;=\; \sum_{x \in M} \Tr (E_x P)\:.
\eeq
The expression~$\Tr (E_x P)$ can be understood as the localization of the trace
at the space-time point~$x$, and summing over all space-time points
gives the total trace.
When forming more complicated composite expressions, it is convenient to use
the short notations
\beq \label{notation}
P(x,y) \;=\; E_x\,P\,E_y \spc {\mbox{and}} \spc
\Psi(x) \;=\; E_x\,\Psi \:.
\eeq
The operator~$P(x,y)$ maps~$E_y(H) \subset H$ to~$E_x(H)$, and it is often
useful to regard it as a mapping only between these subspaces,
\[ P(x,y)\;:\; E_y(H) \: \rightarrow\: E_x(H)\:. \]
Using~(\ref{oc}), we can write the vector~$P \Psi$ as follows,
\[ (P\Psi)(x) \;=\; E_x\: P \Psi \;=\; \sum_{y \in M} E_x\,P\,E_y\:\Psi
\;=\; \sum_{y \in M} (E_x\,P\,E_y)\:(E_y\,\Psi) \:, \]
and thus
\beq \label{diskernel}
(P\Psi)(x) \;=\; \sum_{y \in M} P(x,y)\: \Psi(y)\:.
\eeq
This relation resembles the representation of an operator with an integral kernel.
Therefore, we call~$P(x,y)$ the {\em{discrete kernel}} of the fermionic projector.

We can now set up our variational principle. We define the
{\em{closed chain}}~$A_{xy}$ by
\beq \label{cc}
A_{xy} \;=\; P(x,y)\: P(y,x) \;=\; E_x \:P\: E_y \:P\: E_x \:;
\eeq
it maps~$E_x(H)$ to itself.
Let~$\lambda_1,\ldots,\lambda_{2n}$ be the zeros of the characteristic polynomial
of~$A_{xy}$, counted with multiplicities. We define the {\em{spectral weight}}~$|A_{xy}|$
by
\[ |A_{xy}| \;=\; \sum_{j=1}^{2n} |\lambda_j|\:. \]
Similarly, one can take the spectral weight of powers of~$A_{xy}$, and by summing
over the space-time points we get positive numbers depending only on the
form of the fermionic projector relative to the space-time projectors.
Our variational principle is to
\beq \label{vary}
{\mbox{minimize}} \quad \sum_{x,y \in M} |A_{xy}^2|
\eeq
by considering variations of the fermionic projector which satisfy the constraint
\beq \label{constraint}
\sum_{x,y \in M} |A_{xy}|^2 = {\mbox{const}} \:.
\eeq
In the variation we also keep the number of particles~$f$ as well as
discrete space-time fixed.
Using the method of Lagrange multipliers,
for every minimizer~$P$ there is a real parameter~$\mu$ such that~$P$
is a stationary point of the {\em{action}}
\beq \label{Sdef}
{\mathcal{S}}_\mu[P] \;=\; \sum_{x,y \in M} {\mathcal{L}}_\mu[A_{xy}]
\eeq
with the {\em{Lagrangian}}
\beq \label{Ldef}
{\mathcal{L}}_\mu[A] \;=\; |A^2| - \mu\: |A|^2 \:.
\eeq

This variational principle was first introduced in~\cite{PFP}. In~\cite{F1}
it is analyzed mathematically, and it is shown in particular that
minimizers exist:
\begin{Thm} \label{thmn1}
The variational principle~(\ref{vary}, \ref{constraint}) attains its minimum.
\end{Thm}

\section{Discussion, the Underlying Physical Principles} \label{sec2}
\markboth{Felix Finster}{The Principle of the Fermionic Projector}
We come to the physical discussion. Obviously, our mathematical framework does not
refer to an underlying space-time continuum, and our variational principle is set up
intrinsically in discrete space-time. In other words, our approach is {\em{background free}}.
Furthermore, the following physical principles are respected, in a sense we briefly explain.
\begin{itemize}
\item The {\bf{Pauli Exclusion Principle}}:
We interpret the vectors in the image of~$P$ as the quantum mechanical states of the particles
of our system. Thus, choosing a basis~$\Psi_1,\ldots, \Psi_f
\in P(H)$, the~$\Psi_i$ can be thought of as the wave functions of the occupied states
of the system.
Every vector $\Psi \in H$
either lies in the image of $P$ or it does not.
Via these two conditions, the fermionic projector encodes for every state
$\Psi$ the occupation numbers $1$ and $0$, respectively, but it is
impossible to describe higher occupation numbers.
More technically, we can form the anti-symmetric many-particle wave function
\[ \Psi \;=\; \Psi_1 \wedge \cdots \wedge \Psi_f \:. \]
Due to the anti-symmetrization, this definition of~$\Psi$ is (up to a phase)
independent of the choice of the basis $\Psi_1,\ldots, \Psi_f$.
In this way, we can associate to every fermionic projector a fermionic
many-particle wave function which obeys the Pauli Exclusion Principle.
For a detailed discussion we refer to~\cite[\S3.2]{PFP}.

\item A {\bf{local gauge principle}}:
Exactly as in Hilbert spaces, a linear operator~$U$ in~$H$ is called
{\em{unitary}} if
\[ \bra U \Psi \:|\: U \Phi \ket \;=\; \bra \Psi \:|\: \Phi \ket
\spc {\mbox{for all $\Psi, \Phi \in H$}}. \]
It is a simple observation that a joint unitary transformation
of all projectors,
\beq \label{unit}
E_x \;\to\; U E_x U^{-1} \:, \qquad
P \;\to\; U P U^{-1} \spc {\mbox{with~$U$
unitary}}
\eeq
keeps our action~(\ref{vary}) as well as the constraint~(\ref{constraint})
unchanged, because
\begin{eqnarray*}
P(x,y) &\to& U\:P(x,y)\:U^{-1} \:,\spc
A_{xy} \;\to\; U A_{xy} U^{-1} \\
\det (A_{xy} - \lambda \1) &\to&
\det \!\left( U (A_{xy} - \lambda\1)\: U^{-1} \right) \;=\;
\det (A_{xy} - \lambda\1)\:,
\end{eqnarray*}
and so the~$\lambda_j$ stay the same. Such unitary transformations can
be used to vary the fermionic projector. However, since we want to keep discrete
space-time fixed, we are only allowed to consider unitary
transformations which do not change the space-time projectors,
\beq \label{gauge1}
E_x \;=\; U E_x U^{-1} \spc {\mbox{for all $x \in M$}}\:.
\eeq
Then~(\ref{unit}) reduces to the transformation of the fermionic projector
\beq \label{gauge2}
P \;\to\; U P U^{-1}\:.
\eeq
The conditions~(\ref{gauge1}) mean that~$U$ maps every subspace~$E_x(H)$ into
itself. Hence~$U$ splits into a direct sum of unitary transformations
\beq \label{local}
U(x) \;:=\; U E_x \;:\; E_x(H) \: \rightarrow\: E_x(H) \:,
\eeq
which act ``locally'' on the subspaces associated to the individual space-time
points.

Unitary transformations of the form~(\ref{gauge1}, \ref{gauge2}) can be identified
with local gauge transformations. Namely, using the notation~(\ref{notation}),
such a unitary transformation~$U$ acts on a vector~$\Psi \in H$ as
\[ \Psi(x) \;\longrightarrow\; U(x)\: \Psi(x)\:. \]
This formula coincides with the well-known transformation law of wave functions
under local gauge transformations (for more details see~\cite[\S1.5 and \S3.1]{PFP}).
We refer to the group of all unitary transformations of the form~(\ref{gauge1}, \ref{gauge2})
as the {\em{gauge group}}. The above argument shows that our variational
principle is {\em{gauge invariant}}.
Localizing the gauge transformations according to~(\ref{local}), we obtain
at any space-time point~$x$ the so-called {\em{local gauge group}}.
The local gauge group is the group of
isometries of~$E_x(H)$ and can thus be identified with the group~$U(n,n)$.
Note that in our setting the local gauge group cannot be chosen arbitrarily,
but it is completely determined by the spin dimension.
\item The {\bf{equivalence principle}}:
At first sight it might seem impossible to speak of the equivalence principle without
having the usual space-time continuum. What we mean is the following more general notion.
The equivalence principle can be expressed by the invariance of the physical equations
under general coordinate transformations. In our setting, it makes no sense to speak
of coordinate transformations nor of the diffeomorphism group because we have no
topology on the space-time points. But instead, we can take the largest group which can act
on the space-time points: the group of all permutations of~$M$. Our variational principle
is obviously {\em{invariant under the permutation group}} because permuting the space-time
points merely corresponds to reordering the summands in~(\ref{vary}, \ref{constraint}).
Since on a Lorentzian manifold, every diffeomorphism is bijective and can thus be regarded
as a permutation of the space-time points, the invariance of our variational principle
under the permutation group can be considered as a generalization of the equivalence principle.
\end{itemize}
An immediate objection to the last paragraph is that the symmetry under permutations of the
space-time points is not compatible with the topological and causal structure of a Lorentzian
manifold, and this leads us to the discussion of the physical principles which are
{\em{not}} taken into account in our framework. Our definitions involve
{\bf{no locality}} and {\bf{no causality}}. We do not consider these principles as being
fundamental. Instead, our concept is that the causal structure is induced on the space-time
points by the minimizer~$P$ of our variational principle. In particular, minimizers should
spontaneously break the above permutation symmetry to a smaller symmetry group, which, in a certain limiting case describing the vacuum, should reduce to Poincar{\'e} invariance. Explaining in detail how
this is supposed to work goes beyond the scope of this short article
(for a first step in the mathematical analysis of spontaneous symmetry
breaking see~\cite{F2}). In order to tell
the reader right away what we have in mind, we shall first simply assume the causal
structure of Minkowski space and consider our action in the setting of relativistic
quantum mechanics (Section~\ref{sec3}). This naive procedure will {\em{not}} work, but it
will nevertheless illustrate our variational principle and reveal a basic difficulty.
In Section~\ref{sec4} we will then outline the connection to the continuum theory as
worked out in~\cite{PFP}.

\section{Naive Correspondence to a Continuum Theory} \label{sec3}
\markboth{Felix Finster}{The Principle of the Fermionic Projector}
Let us see what happens if we try to get a connection between the framework of
Section~\ref{sec1} and relativistic quantum mechanics in the simplest possible way.
To this end, we just replace~$M$ by the space-time continuum~$\R^4$ and the sums
over~$M$ by space-time integrals. For a vector~$\Psi \in H$, the corresponding
$\Psi(x) \in E_x(H)$ as defined by~(\ref{notation}) should be a 4-component Dirac
wave function, and the scalar product $\bra \Psi(x) \,|\, \Phi(x) \ket$ on~$E_x(H)$
should correspond to the usual Lorentz invariant scalar product on Dirac
spinors~$\overline{\Psi} \Phi$
with~$\overline{\Psi} = \Psi^\dagger \gamma^0$ the adjoint spinor. Since this last
scalar product is indefinite of signature~$(2,2)$, we are led to choosing~$n=2$, so that
the spin dimension is~$(2,2)$.

In view of~(\ref{diskernel}), the discrete kernel should in the continuum go over to the
integral kernel of an operator~$P$ on the Dirac wave functions,
\[ (P \Psi)(x) \;=\; \int_M P(x,y)\, \Psi(y)\: d^4y \:. \]
The image of~$P$ should be spanned by the occupied fermionic states. We take Dirac's
concept literally that in the vacuum all negative-energy states are occupied by fermions
forming the so-called {\em{Dirac sea}}. This leads us to describe the vacuum by the integral
over the lower mass shell
\beq \label{Fourier}
P(x,y) \;=\; \int \frac{d^4k}{(2 \pi)^4}\: (k \slsh+m)\:
\delta(k^2-m^2)\: \Theta(-k^0)\: e^{-ik(x-y)}
\eeq
(we consider for simplicity only one Dirac sea of mass~$m$; the factor
$(k \slsh+m)$ is needed in order to satisfy the Dirac equation
$(i \Pdd_x - m)\, P(x,y) = 0$).

We now consider our action for the above fermionic projector. Since
we do not want to compute the Fourier integral~(\ref{Fourier}) in detail, we simply choose
$x$ and $y$ for which the integrals in~(\ref{Fourier}) exist (for details see below)
and see what we get using only the {\em{Lorentz symmetry}} of~$P$. We can clearly write
$P(x,y)$ as
\[ P(x,y) \;=\; \alpha\, (y-x)_j \gamma^j + \beta\:\1 \]
with two complex parameters~$\alpha$ and~$\beta$. Taking the complex conjugate
of~(\ref{Fourier}), we see that
\[ P(y,x) \;=\; \overline{\alpha}\, (y-x)_j \gamma^j + \overline{\beta}\:\1 \:. \]
As a consequence,
\beq \label{1}
A_{xy} \;=\; P(x,y)\, P(y,x) \;=\; a\, (y-x)_j \gamma^j + b\, \1
\eeq
with real parameters~$a$ and $b$ given by
\beq \label{ab}
a \;=\; \alpha \overline{\beta} + \beta \overline{\alpha} \:,\spc
b \;=\; |\alpha|^2 \,(y-x)^2 + |\beta|^2 \:.
\eeq
Using the formula~$(A_{xy} - b \1)^2 = a^2\:(y-x)^2$, one can easily compute the zeros
of the characteristic polynomial of~$A_{xy}$,
\[ \lambda_1 \;=\; \lambda_2 \;=\; b + \sqrt{a^2\: (y-x)^2} \:,\spc
\lambda_3 \;=\; \lambda_4 \;=\; b - \sqrt{a^2\: (y-x)^2}\:. \]
If the vector~$(y-x)$ is spacelike, we conclude from the inequality $(y-x)^2<0$
that the argument of the above square root
is negative. As a consequence, the~$\lambda_j$ appear in {\em{complex conjugate pairs}},
\[ \overline{\lambda_1} \;=\; \lambda_3 \:,\spc
\overline{\lambda_2} \;=\; \lambda_4\:. \]
Furthermore, the~$\lambda_j$ all have the same absolute value $|\lambda_j|=:|\lambda|$,
and thus the Lagrangian~(\ref{Ldef}) reduces to
\[ {\mathcal{L}}_\mu[A] \;=\; |\lambda|^2 \left(4 - 16\, \mu \right) . \]
This simplifies further if we choose the Lagrange multiplier equal
to~$\frac{1}{4}$, because then the action vanishes identically.
If conversely~$(y-x)$ is timelike, the~$\lambda_i$ are all real. Using~(\ref{ab}),
one easily verifies that they are all {\em{positive}} and
thus~${\mathcal{L}}_{\frac{1}{4}}[A] = (\lambda_1-\lambda_3)^2$.
We conclude that
\beq \label{as}
 {\mathcal{L}}_{\frac{1}{4}}[A_{xy}] \;=\; \left\{ \begin{array}{cl} 4 a^2\: (y-x)^2 &
{\mbox{if $(y-x)$ is timelike}} \\
0 & {\mbox{if $(y-x)$ is spacelike}}\:.
\end{array} \right.
\eeq

This consideration gives a simple {\em{connection to causality}}:
In the two cases where~$(y-x)$ is timelike or spacelike, the spectral properties
of the matrix~$A_{xy}$ are completely different (namely, the~$\lambda_j$ are real
or appear in complex conjugate pairs, respectively), and this leads to a completely
different form of the action~(\ref{as}). More specifically, if the~$\lambda_j$ are
non-real, this property is (by continuity) preserved under small perturbations of~$A_{xy}$.
Thinking of a dynamical situation, this suggests that perturbations of~$P(x,y)$
for spacelike $(y-x)$ should not effect the action or, in other words, that
events at points $x$ and $y$ with spacelike separation should not be related to each
other by our variational principle. We remark that choosing~$\mu=\frac{1}{4}$ is
justified by considering the Euler-Lagrange equations corresponding to our variational
principle, and this also makes the connection to causality clearer
(see~\cite[\S3.5 and \S5]{PFP}).

Apart from the oversimplifications and many special assumptions, the main flaw
of this section is that the Fourier integral~(\ref{Fourier})
does not exist for all $x$ and $y$. More precisely, $P(x,y)$ is a well-defined
distribution, which is even a smooth function if~$(y-x)^2 \neq 0$.
But {\em{on the light cone}}
$(y-x)^2=0$, this distribution is {\em{singular}} (for more
details see~\cite[\S2.5]{PFP}). Thus on the light cone, the pointwise product
in~(\ref{1}) is ill-defined and our above arguments fail. The resolution of this problem will be outlined in the next section.

\section{The Continuum Limit} \label{sec4}
\markboth{Felix Finster}{The Principle of the Fermionic Projector}
We now return to the discrete setting of Section~\ref{sec1} and shall explain
how to get a rigorous connection to the continuum theory. One approach is to
study the minimizers in discrete space-time and to try to recover structures
known from the continuum. For example, in view of the spectral properties
of~$A_{xy}$ in Minkowski space as discussed in the previous section, it is
tempting to introduce in discrete space-time the following notion
(this definition is indeed symmetric in~$x$ and~$y$,
see~\cite[\S3.5]{PFP}).
\begin{Def} \label{def41}
Two discrete space-time points~$x,y \in M$ are called
{\bf{timelike}} separated if the zeros~$\lambda_j$ of the
characteristic polynomial of~$A_{xy}$ are all real.
They are said to be {\bf{spacelike}} separated if
the~$\lambda_j$ are all non-real and have the same absolute value.
\end{Def}
The conjecture is that if the number of space-time points and the number of particles both
tend to infinity at a certain relative rate, the above ``discrete causal structure''
should go over to the causal structure of a Lorentzian manifold.
Proving this conjecture under suitable assumptions is certainly a challenge.
But since we have a precise mathematical framework in discrete space-time,
this seems an interesting research program.

Unfortunately, so far not much work has been done on the discrete models, and at present
almost nothing is known about the minimizers in discrete space-time. For this reason,
there seems no better method at the moment than to impose that the fermionic projector
of the {\em{vacuum}} is obtained from a Dirac sea configuration by a suitable regularization
process on the Planck scale~\cite[Chapter~4]{PFP}. Since we do not know how the
physical fermionic projector looks like on the Planck scale, we use the {\em{method of
variable regularization}} and consider a large class of regularizations~\cite[\S4.1]{PFP}.

When introducing the fermionic projector of the vacuum, we clearly put in the causal
structure of Minkowski space as well as the free Dirac equation ad hoc. What makes the
method interesting is that we then introduce a {\em{general interaction}} by inserting
a general (possibly nonlocal) perturbation operator into the Dirac equation. Using methods of hyperbolic PDEs (the so-called {\em{light-cone
expansion}}), one can describe the fermionic projector with interaction
in detail~\cite[\S2.5]{PFP}. It turns out that the regularization of the fermionic
projector with interaction is completely determined by the regularization of the
vacuum (see~\cite[\S4.5 and Appendix~D]{PFP}). Due to the regularization, the singularities
of the fermionic projector have disappeared, and one can consider the
{\em{Euler-Lagrange equations}} corresponding to our variational principle
(see~\cite[\S4.5 and Appendix~F]{PFP}). Analyzing the dependence on the regularization
in detail, we can perform an expansion in powers of the Planck length. This gives
differential equations involving Dirac and gauge fields, which involve a small
number of so-called {\em{regularization parameters}}, which depend on the regularization
and which we treat as free parameters (see~\cite[\S4.5 and Appendix~E]{PFP}).
This procedure for analyzing the Euler-Lagrange equations in the continuum is called
{\em{continuum limit}}. We point out that only the singular behavior of~$P(x,y)$
on the light cone enters the continuum limit, and this gives causality.

\section{Obtained Results} \label{sec5}
\markboth{Felix Finster}{The Principle of the Fermionic Projector}
In~\cite[Chapters~6-8]{PFP} the continuum limit is analyzed in spin dimension~$(16,16)$
for a fermionic projector of the vacuum, which is
the direct sum of seven identical massive sectors and one massless
left-handed sector, each of which is composed of
three Dirac seas. Considering general chiral and (pseudo)scalar potentials, we find
that the sectors spontaneously form pairs, which are referred to
as {\em{blocks}}. The resulting {\em{effective interaction}} can be described
by chiral potentials corresponding to the effective gauge group
\[ SU(2) \times SU(3) \times U(1)^3 \;. \]
This model has striking similarity to the standard model if the block
containing the left-handed sector is identified with the leptons and the
three other blocks with the quarks. Namely, the effective gauge fields
have the following properties.
\begin{itemize}
\item The $SU(3)$ corresponds to an unbroken gauge symmetry. The
$SU(3)$ gauge fields couple to the quarks exactly
as the strong gauge fields in the standard model.
\item The $SU(2)$ potentials are left-handed and couple to the leptons
and quarks exactly as the weak gauge potentials in the standard model.
Similar to the CKM mixing in the standard model,
the off-diagonal components of these potentials must involve a
non-trivial mixing of the generations.
The $SU(2)$ gauge symmetry is spontaneously broken.
\item The $U(1)$ of electrodynamics can be identified with an Abelian
subgroup of the effective gauge group.
\end{itemize}
The effective gauge group is larger than the gauge group of the standard
model, but this is not inconsistent because a more detailed analysis of
our variational principle should give further constraints for the
Abelian gauge potentials. Moreover, there are the following differences
to the standard model, which we derive mathematically without working
out their physical implications.
\begin{itemize}
\item The $SU(2)$ gauge field tensor $F$ must be simple in the sense
that $F=\Lambda \:s$ for a real 2-form $\Lambda$ and an $su(2)$-valued
function $s$.
\item In the lepton block, the off-diagonal $SU(2)$ gauge potentials are
associated with a new type of potential, called nil potential, which couples
to the right-handed component.
\end{itemize}

\section{Outlook: The Classical Gravitational Field} \label{sec6}
\markboth{Felix Finster}{The Principle of the Fermionic Projector}
The permutation symmetry of our variational principle
as discussed in Section~\ref{sec2} guarantees that the equations
obtained in the continuum limit are invariant under diffeomorphisms.
This gives us the hope that classical gravity might already be taken
into account, and that even quantum gravity might be incorporated in
our framework if our variational principle is studied beyond
the continuum limit. Unfortunately, so far these questions have
hardly been investigated. Therefore, at this point we
leave rigorous mathematics and must enter the realm of what
a cricial scientist might call pure speculation.
Nevertheless, the following discussion might be helpful to give
an idea of what our approach is about, and it might also
give inspiration for future work in this area.

The only calculations for gravitational fields carried out so far
are the calculations for linearized gravity~\cite[Appendix~B]{F3}.
The following discussion of classical gravity is based on these calculations.
For the metric, we consider a linear perturbation $h_{jk}$ of the
Minkowski metric
$\eta_{jk}={\mbox{diag}}(1,-1,-1,-1)$,
\[ g_{jk}(x) \;=\; \eta_{jk} \:+\: h_{jk}(x) \: . \]
In linearized gravity, the diffeomorphism invariance corresponds
to a large freedom to transform the~$h_{jk}$ without changing
the space-time geometry (this freedom is usually referred to as ``gauge freedom'', but we point out for clarity that it is not related
to the ``local gauge freedom'' as discussed in Section~\ref{sec2}).
This freedom can be used to arrange that (see e.g.\ \cite{LL})
\[ \partial^k h_{jk} \;=\; \frac{1}{2} \:\partial_j
h_{kl}\: \eta^{kl}\:. \]
Computing the corresponding Dirac operator and performing the
light-cone expansion, the first-order perturbation of the fermionic projector
takes the form
\begin{eqnarray}
\lefteqn{ \Delta{P}(x,y) \;=\; {\mathcal{O}}(\xi\slsh\, z^{-1})
+ {\mathcal{O}}(\xi^{[k} \gamma^{l]} \, z^{-1})
+ {\mathcal{O}}(m) + {\mathcal{O}}((h_{ij})^2)} \label{lce0} \\
&&\!\!\! \left. \begin{array}{l} \displaystyle
        +\frac{1}{2}
        \left( \int_x^y h^k_j \right) \xi^j \frac{\partial}{\partial y^k}
        \;P(x,y) \\[1em]
\displaystyle -\frac{i}{16 \pi^3} \:
        \left( \int_x^y (2 \alpha -1) \; \gamma^i \: \xi^j \:\xi^k
        \; (h_{jk},_i - h_{ik},_j)  \right) d\alpha \:z^{-2} \\[1em]
\displaystyle -\frac{1}{32 \pi^3} \:
        \left( \int_x^y \varepsilon^{ijlm} \; (h_{jk},_i
        - h_{ik},_j) \: \xi^k \; \xi_l \: \rho \gamma_m \right)
        d\alpha \:z^{-2}
\end{array} \right\} \label{lce1} \\
&&+\frac{i}{32 \pi^3} \: \left( \int_x^y (\alpha^2 - \alpha) \; \xi^j \:
        \gamma^k \: G_{jk} \right) d\alpha \:z^{-1}\:, \label{lce2}
\end{eqnarray}
where we set~$\xi \equiv y-x$, the integrals go along
straight lines joining the points~$x$ and~$y$,
\[ \int_x^y f \:d\alpha \;=\;
\int_0^1 f(\alpha y + (1-\alpha) x)\, d\alpha\:, \]
and~$z^{-1}, z^{-2}$ are distributions
which are singular on the light cone,
\[ z^{-1} \;=\; \frac{\mbox{PP}}{\xi^2} \:+\: i \pi \delta (\xi^2) \:
\epsilon (\xi^0) \:,\quad
z^{-2} \;=\; \frac{\mbox{PP}}{\xi^4} \:-\: i \pi \delta'(\xi^2) \:
\epsilon (\xi^0) \]
(where~PP denotes the principal part, and~$\epsilon$ is the
step function~$\epsilon(x)=1$ if~$x>0$ and~$\epsilon(x)=-1$ otherwise).
In this formula we only considered the most singular contributions
on the light cone and did no take into account the higher orders
in the rest mass~$m$ of the Dirac particles (for details
see~\cite{F3}). Nevertheless, the above formula gives us some general
information on how the fermionic projector depends on a classical
gravitational field.
The contribution~(\ref{lce1}) describes an ``infinitesimal
deformation'' of the light cone corresponding to the fact that
the gravitational field affects the causal structure.
Since it involves at most first derivatives of the metric, the
curvature does not enter, and thus~(\ref{lce1}) can be compensated
by a gauge and an infinitesimal coordinate tranformation.
The diffeomorphism invariance of the equations of the continuum
limit ensures that the contribution~(\ref{lce1}) drops out of these
equations (and this can also be verified by a direct computation
of the closed chain). We conclude that the equations of the continnum limit
will be governed by the contribution~(\ref{lce2}). It is remarkable
that the {\em{Einstein tensor}} $G_{jk}$ appears.
Thus, provided that the equations of the continuum limit give
sufficiently strong constraints, we obtain the vacuum Einstein equations.

The situation becomes even more interesting if fermionic matter is
involved. In this case, the wave function~$\Psi$ of a particle
(or similarly anti-particle) will lead to a perturbation of the
fermionic projector of the form
\beq \label{matter}
\Delta{P}(x,y) \;=\; - \Psi(x)\, \overline{\Psi(y)}\:.
\eeq
Performing a multi-pole expansion around~$x$, the
zeroth moment~$-\Psi(x) \overline{\Psi(x)}$
corresponds to the electromagnetic current and should
be taken care of by the Maxwell equations. The first moment
\[ -(y-x)^j\;\Psi(x)\, \partial_j \overline{\Psi(x)} \]
is proportional to the energy-momentum tensor of the Dirac wave function.
Imposing that this contribution should be compensated by the
first moment of~(\ref{lce2}), we obtain a relation of the form
\beq \label{relnaive}
\frac{i}{32 \pi^3} \: \frac{1}{6} \;\xi^j \: G_{jk} \;z^{-1}
\;=\; \xi^j\: T_{jk}[\Psi] \:.
\eeq
This calculation was too naive, because the left side of the equation
involves the singular distribution~$z^{-1}$, whereas the right side
is smooth. This is also the reason why the method of the continuum
limit as developed in~\cite{PFP} cannot be applied directly to the
gravitational field. On the other hand, this is not to be expected,
because the formalism of the continuum limit only gives
dimensionless constants, whereas the gravitational constant
has the dimension of length squared. These extra length dimensions enter~(\ref{relnaive}) by the factor~$z^{-1}$.
The simplest method to convert~(\ref{relnaive}) into a reasonable
differential equation is to argue that the concept of the
space-time continuum should be valid only down to the Planck scale,
where the discreteness of space-time should lead to some kind of
``ultraviolet regularization.'' Thus it seems natural to replace
the singular factor~$z^{-1}$ by the value of this factor on the
Planck scale. This leads to an equation of the form
\beq \label{relright}
G_{jk}\; \frac{1}{l_P^2} \;\sim\; T_{jk} \:,
\eeq
where~$l_P$ denotes the Planck length. These are the precisely the
Einstein equations. We point out that the above argument is not
rigorous, in particular because the transition from~(\ref{relnaive})
to~(\ref{relright}) would require a methods which go beyond
the formalism of the continuum limit.
Nevertheless, our consideration seems to explain why the Planck
length enters the Einstein equations and in particular why the coupling
of matter to the gravitational field is so extremely weak.
Also, we get some understanding for how the Einstein equations
could be derived from our variational principle.

\section{Outlook: The Field Quantization} \label{sec7}
\markboth{Felix Finster}{The Principle of the Fermionic Projector}
We hope that in our approach, the field quantization is taken into account
as soon as one goes beyond the continuum limit
and takes into account the discreteness of space-time.
Since the basic mechanism should be the same for the gravitational
field as for any other bosonic field, for simplicity
we can here consider only an electromagnetic field.
The basic ideas are quite old and were one of the motivations for
thinking of the principle of the fermionic projector~\cite{F0}.
Nevertheless, the details have not been worked out in the meantime,
simply because it seemed more important to first get a rigorous
connection to the continuum theory by analyzing the continuum limit.
Thus the following considerations are still on the same speculative
level as nine years ago. In order to convey the reader some of the
spontaneity of the early text, we here simply give a slightly
revised English translation of~\cite[Section~1.4]{F0}.

In preparation of the discussion of field quantization, we want
to work out why quantized bosonic fields are needed,
i.e.\ what the essence of a ``quantization'' of these fields is.
To this aim, we shall analyze to which extent we can get a connection
to quantum field theory by considering {\em{classical}} gauge fields.
For simplicity, we restrict attention to one type of particles and
an electromagnetic interaction, but the considerations apply just
as well to a general interaction including gravitational fields.
Suppose that when describing the interacting system of fermions in
the continuum limit we get the system of coupled differential equations
\beq \label{DM}
(i \Pdd + e \Aslsh - m) \: \Psi \;=\; 0 \:,\qquad
F^{ij}_{\;\;,j} \;=\; e \: \overline{\Psi} \gamma^i \Psi\:.
\eeq
These equations are no longer valid at energies as high as the
Planck energy, because the approximations used in the formalism
of the continuum limit are no longer valid. Our variational principle
in discrete space-time should then still describe our system,
but at the moment we do not know how the corresponding interaction
looks like. For simplicity, we will assume in what follows that the
fermions do not interact at such high energies. In this way, we get
in the classical Maxwell equations a natural cutoff for
very large momenta.

When describing~(\ref{DM}) perturbatively, one gets Feynman diagrams.
To this end we can proceed just as in~\cite{BD}: We expand~$\Psi$
and~$A$ in powers of~$e$,
\[ \Psi \;=\; \sum_{j=0}^\infty e^j \: \Psi^{(j)} \:,\qquad
A \;=\; \sum_{j=0}^\infty e^j \: A^{(j)} \]
and substitute these expansions
in the differential equations~(\ref{DM}). In these
equations, the contributions to every order in~$e$ must vanish,
and thus one solves for the highest appearing index~$^{(j)}$.
In the Lorentz gauge, we thus obtain the formal relations
\beq \label{iter}
\Psi^{(j)} \;=\; - \sum_{k+l=j-1} (i \Pdd -m)^{-1} \: \left(
    \Aslsh^{(k)} \: \Psi^{(l)} \right) \:,\quad
A^{(j)}_i \;=\; - \sum_{k+l=j-1}  \Box^{-1} \: \left(
    \overline{\Psi}^{(k)} \gamma_i \Psi^{(l)} \right) \:,
\eeq
which by iterative substitutions can be brought into a more
explicit form. Taking into account the pair creation, we obtain
additional diagrams which contain closed fermion lines,
due to the Pauli Exclusion Principle with the correct relative signs.
In this way we get all Feynman diagrams.

We come to the renormalization. Since we obtain all the Feynman
diagrams of quantum field theory, the only difference of our approach
to standard quantum field theory is the natural cutoff for large
momenta. In this way all ultraviolet divergences disappear, and the
difference between naked and effective coupling constants becomes
finite. One can (at least in principle) express the effective
coupling constants in terms of the naked coupling constants by
adding up all the contributions by the self-interaction.
Computations using the renormalization group show that the effective
masses and coupling constants depend on the energy. The effective
constants at the Planck scale should be considered as our naked
coupling constants.

The fact that the theory can be renormalized is important for us,
because this ensures that the self-interaction can be described
merely by a change of the masses and coupling constants.
But renormalizability is not absolutely necessary for a meaningful theory.
For example, the renormalizability of diagrams is irrelevant for classes
of diagrams which (with our cutoff) are so small that they are
negligible. Furthermore, one should be aware that the introduction
of a cutoff is an approximation which has no justification.
In order to understand the self-interaction at high energies
one would have to analyze our variational principle without
using the formalism of the continuum limit.

We explained the connection to the Feynman diagrams and the
renormalization in order to point out that perturbative
quantum field theory is obtained already with {\em{classical}}
bosonic fields if one studies the coupled interaction between the
classical field and the fermions. With second quantization of the
gauge fields one can obtain the Feyman diagrams using Wick's theorem
in a more concise way, but at this point it is unnecessary both
from the mathematical and physical point of view to go over from
classical to quantized bosonic fields. In particular, one should
be aware of the fact that all the high precision tests of quantum
field theory (like the Lamb shift and the anomalous $g$ factor) are
actually no test of the field quantization. One does not need to
think of a photon line as an ``exchange of a virtual photon''; the
photon propagator can just as well be considered simply as the
operator~$\Box^{-1}$ in~(\ref{iter}), which appears in the
perturbation expansion of the coupled differential equations~(\ref{DM}).
Also the equation~$E=\hbar \omega$, which in a graphic language
tells us about the ``energy of one photon,'' does not make a statement
on the field quantization. This can be seen as follows: In physics,
the energy appears in two different contexts. In classical field theory,
the energy is a conserved quantity following from the time
translation invariance of the Lagrangian. In quantum theory, on the
other hand, the sum of the frequencies of the wave functions and
potentials is conserved in any interaction, simply because in the
perturbation expansion plane waves of different frequencies are
orthogonal. These ``classical'' and ``quantum mechanical'' energies
are related to each other via the equation~$E=\hbar \omega$.
Planck's constant can be determined without referring to the
electromagnetic field (for example via the Compton wavelength of the
electron). Since the classical and quantum mechanical energies are
both conserved, it is clear that the relation~$E=\hbar \omega$ must
hold in general. (Thus the energy transmitted by a photon line of
frequency~$\omega$ really is~$\hbar \omega$.)

After these considerations there remain only a few effects which
are real tests of the field quantization. More precisely, these are
the following observations,
\begin{enumerate}
\item Planck's radiation law
\item the Casimir effect
\item the wave-particle duality of the electromagnetic field, thus
for example the double-slid experiment
\end{enumerate}
For the derivation of Planck's radiation law, one uses that the
energy of an electromagnetic radiation mode cannot take continuous
values, but that its energy is quantized in steps of~$\hbar \omega$.
The Casimir effect measures the zero point energy of the
radiation mode.
In order to understand field quantization, one needs to find a
convincing explanation for the above observations. However, the
formalism of quantum field theory does not immediately follow from
the above observations. For example, when performing
canonical quantization one assumes that each radiation mode can be described
by a quantum mechanical oscillator. This is a possible explanation,
but it is not a compelling consequence of the discreteness of the
energy levels.

We shall now explain how the above observations could be
explained in the framework of the principle of the fermionic
projector. In order to work out the difference between the
continuum limit and the situation in discrete space-time, we will
discuss several examples. It will always be sufficient to work
also in discrete space-time with the classical notions. For example,
by an electromagnetic wave in discrete space-time we mean a variation
of the fermionic projector which in the continuum limit can be
described via a perturbation of the Dirac operator by a classical
electromagnetic field.

We begin with a simple model in discrete space-time, namely a
completely filled Dirac sea and an electromagnetic field in the form
of a radiation mode. We want to analyze the effect of a variation of
the amplitude of the electromagnetic wave. In the continuum limit,
we can choose the amplitude arbitrarily, because the Maxwell equations
will in any case be satisfied. However, the situation is more
difficult in discrete space-time. Then the variation of the amplitude
corresponds to a variation of the fermionic projector.
However, when performing the perturbation expansion for~$P$ in
discrete space-time, we need to take into account several contributions
which could be left out in the continuum limit. These additional
contributions do not drop out of the Euler-Lagrange equations corresponding
to our variational principle. If these equations are satisfied for
a given fermionic projector~$P$, we cannot expect that they will
still hold after changing the amplitude of the electromagnetic wave.
More generally, in discrete space-time there seems to be no
continuous family~$P(\tau)$ of solutions of the Euler-Lagrange equations.
This means in particular that the amplitude of the electromagnetic wave
can take only discrete values.

Alternatively, the difference between the continuum limit and the
description in discrete space-time can be understood as follows:
In discrete space-time, the number~$f$ of particles is an integer.
If for different values of~$f$ we construct a fermionic projector
of the above form, the amplitude of the corresponding electromagnetic
wave will in general be different. Let us assume for simplicity that
for each~$f$ (in a reasonable range) there is exactly one such
projector~$P_f$ with corresponding amplitude~$A_f$. Since~$f$ is
not known, we can choose~$f$ arbitrarily. Thus the amplitude
of the wave can take values in the discrete set~$\{A_f\}$.
In the continuum limit, however, the fermionic projector is an
operator of infinite rank. Thus it is clear that now we do not get
a restriction for the amplitude of the electromagnetic wave, and
the amplitude can be varied continuously.

We conclude that in discrete space-time a natural ``quantization'' of
the amplitude of the electromagnetic wave should appear. Before we can get a connection to the Planck radiation and the
Casimir effect, we need to refine our consideration. Namely, it seems
unrealistic to consider an electromagnetic wave which is spread
over the whole of space-time. Thus we now consider a wave in
a four-dimensional box (for example with fixed boundary values).
Let us assume that the box has length~$L$ in the spatial directions
and~$T$ in the time direction. In this case, again only discrete
values for the amplitude of the wave should be admissible.
But now the quantization levels should depend on the size of the box,
in particular on the parameter~$T$. Qualitatively, one can expect
that for smaller~$T$ the amplitude of the wave must be larger in order
to perturb the fermionic projector in a comparable way. This means that
the quantization levels become finer if~$T$ becomes larger.
Via the classical energy density of the electromagnetic field,
the admissible amplitudes~$\{A_j\}$ can be translated into
field energies of the wave. Physically speaking, we create a wave
at time~$t$ and annihilate it a later time~$t+T$. Since, according
to our above consideration, the relation~$E=\hbar \omega$ should hold
in any interacting system, we find that the field energy must be
``quantized'' in steps of~$\hbar \omega$. On the other hand, we just
saw that the quantization levels depend on~$T$. In order to avoid
inconsistencies, we must choose~$T$ such that the quantization
steps for the field energy are just~$\hbar \omega$.

In this way we obtain a condition which at first sight seems
very strange: If we generate an electromagnetic wave at some time~$t$,
we must annihilate it at some later time~$t+T$. Such an additional
condition which has no correspondence in the continuum limit,
is called a {\em{non-local quantum condition}}. We derived it
under the assumption of a ``quantization'' of the amplitude
from the equations of the continuum limit (classical field
equations, description of the interaction by Feyman diagrams).
Since the Euler-Lagrange equations of discrete space-time should
in the continuum limit go over to the classical equations, a
solution in discrete space-time should automatically satisfy
the non-local quantum condition.

Of course, the just-derived condition makes no physical sense.
But our system of one radiation mode is also oversimplified.
Thus before drawing further conclusions, let us consider the
situation in more realistic situations: In a system with several
radiation modes, we cannot (in contrast to the situation with
canonical quantization) treat the different modes as being
independent, because the variation of the amplitude of one mode
will influence the quantization levels of all the other
radiation modes. This mutual influence is non-local. Thus an
electromagnetic wave also changes the energy levels of waves which
are in large spacelike distance. The situation becomes even more
complicated if fermions are brought into the system, because then the
corresponding Dirac currents will also affect the energy levels
of the radiation modes.
The complexity of this situation has two consequences: First, we
can make practically no stament on the energy levels, we only know
that the quantization steps are~$\hbar \omega$. Thus we can describe
the energy of the lowest level only statistically. It seems
reasonabl to assume that they are evenly distributed in the interval~$[0, \hbar \omega)$.
Then we obtain for the possible energy levels of each radiation mode
on average the values~$(n+\frac{1}{2}) \hbar \omega$.
Secondly, the non-local quantum conditions are now so complicated
that we can no longer specify them. But it seems well possible that
they can be satisfied in a realistic physical situation. We have the
conception that such non-local quantum conditions determine all
what in the usual statistical interpretation of quantum mechanics
is said to be ``undetermined'' or ``happens by chance''.
We will soon come back to this point when discussing the
wave-particle dualism.

After these consideration we can explain the above observations 1.\
and 2.: Since the energy of each radiation mode is quantized in
steps of~$\hbar \omega$, we obtain Planck's radiation law, whereas
the average energy of~$\frac{1}{2}\, \hbar \omega$ of the ground
state energy explains the Casimir effect. We conclude that under
the assumption of a ``quantization'' of the amplitude of the
electromagnetic wave we come to the same conclusions as with
canonical quantization. The reason is that with the Feynman diagrams
and the equation~$E=\hbar \omega$ we had all the formulas for the
quantitative description at our disposal, and therefore it was
sufficient to work with a very general ``discreteness'' of the energy
levels.

We come to the wave-particle dualism. Since this is a basic effect
in quantum mechanics, which appears similarly for bosons and fermions,
we want to discuss this point in detail. First we want to compare
our concept of bosons and fermions. Obviously, we describe bosons
and fermions in a very different way: the wave functions of the fermions
span the image of the projector~$P$, whereas the bosons correspond (as
described above) to the discrete excitation levels of the classical bosonic fields. In our description, the Fock space or an equivalent formalism
does not appear. It might not seem satisfying that in this way the
analogy in the usual description of bosons and fermions, namely
the mere replacements of commutators by anti-commutators, gets lost.
However, we point out that the {\em{elementary}} bosons and fermions differ
not only by their statistics but also in the following important point.
For the fermions (leptons, quarks) we have a conservation law
(lepton number, baryon number), not so for the gauge bosons.
This difference is taken into account in our formalism: Every fermion
corresponds to a vector in~$P(H)$. We can transform fermions into
each other and can create/annihilate them in pairs. But we cannot
change the total number~$f$ of particles. In particular, we cannot
annihilate a single fermion. In contrast, since the gauge bosons
merely correspond to discrete values of the bosonic fields. They
can be generated or annihilated arbitrarily in the interaction,
provided that the conservation law for energy and momentum is satisfied.

In order to clarify the connection to the Fock space, we briefly mention
how we describe composite particles (for example mesons or baryons).
They are all composed of the elementary fermions. Thus a particle
composed of~$p$ components corresponds to a vector of~$(P(H))^p$.
This representation is not suitable for practical purposes.
It is more convenient to use for the elemenatry fermions the
Fock space formalism. Then the creation/annihlation operators for the
composite particle are a product of~$p$ fermionic creation/annihilation
operators. If~$p$ is even (or odd), we can generate with these
creation/annihilation operators the whole bosonic (or fermionic)
Fock space. In tis way, we obtain for composite particles the
usual formalism. However, we point out that in our description
this formalism has no fundamental significance.

Due to the different treatment of elementary fermions and bosons,
we need to find an explanation for the wave-particle dualism which
is independent of the particular description of these particles.
For a fermion, this is a vector~$\Psi \in P(H)$, for a boson
the gauge field. Thus in any case, the physical object is not the
pointlike particle, but the wave. At first sight this does not seem
reasonable, because we have not at all taken into account the particle
character. Our concept is that the particle character is a consequence
of a ``discreteness'' of the interaction described by our variational
principle.
In order to specify what we mean by ``discreteness'' of the interaction,
we consider the double slid experiment. We work with an electron, but
the consideration applies just as well to a photon, if the wave function
of the electron is replaced by the electromagnetic field.
When it hits the photographic material on the screen, the electron interacts
with the silver atoms, and the film is exposed. In the continuum limit
we obtain the same situation as in wave mechanics: the waves
originating at the two slids are superposed and generate on the
screen an interference pattern.
Similar to our discussion of the electromagnetic radiation mode,
the continuum limit should describe the physical situation only
approximately. But when considerung the variational principle in
discrete space-time, the situation becomes much more complicated.
Let us assume that the interaction in discrete space-time is ``discrete''
in the sense that the electron prefers to interact with only one
atom of the screen. This assumption is already plausible in the
continuum limit. Namely, if the electron interacts with a silver atom,
one electron from the atom must be excited. Since this requires a
certain minimal energy, the kinetic energy of the electron hitting
the screen can excite only a small number of atoms. Thus the interaction
between electron and the screen can take place only at individual
silver atoms; the electron cannot pass its energy continuously
onto the screen.

Under this assumption we get on the screen an exposed dot, and thus
we get the impression of a pointlike particle. At which point of
the screen the interactin takes place is determined by the
detailed form of the fermionic projector~$P$ in discrete space-time.
With the notion introduced above, we can also say that which silver
atom is exposed is determined by non-local quantum conditions.
At this point, the non-locality and non-causality of our
variational principle in discrete space-time becomes important.
Since the non-local quantum conditions are so complicated, we cannot
predict at which point of the screen the interaction will take place.
Even if we repeat the same experiment under seemingly identical
conditions, the global situation will be different. As a consequence,
we can only make statistical statements on the measurements.
From the known continuum limit we know that the probabilities
for the measurements are given by the usual
quantum mechanical expectation values.

At this point we want to close the discussion.
We conclude that the principle of the fermionic projector raises quite
fundamental questions on the structure of space-time, the nature
of field quantization and the interpretation of quantum mechanics.
Besides working out the continuum limit in more detail, it will be a
major goal of future work to give specific answers to these questions.

\end{document}